\documentclass[12pt]{article}
\usepackage{amssymb} 
\usepackage{graphics} 
\usepackage{cite}
\usepackage{epsfig}
\usepackage{color}
\newcommand  {\etal}     {{\it et al.}}

\newcommand  {\Bioch}    {{\it Biochemistry\ }}
\newcommand  {\Biomet}   {{\it Biometrika\ }}
\newcommand  {\Biopol}   {{\it Biopolymers\ }}

\newcommand  {\BJ}       {{\it Biophys.~J.\ }}

\newcommand  {\EL}       {{\it Europhys.\ Lett.\ }}
\newcommand  {\FD}       {{\it Fold.\ Des.\ }}

\newcommand  {\JBP}      {{\it J.\ Biol.\ Phys.\ \ }}

\newcommand  {\JCP}      {{\it J.\ Chem.\ Phys.\ }}
\newcommand  {\JMB}      {{\it J.\ Mol.\ Biol.\ }}

\newcommand  {\Nat}      {{\it Nature\ }}
\newcommand  {\NSB}      {{\it Nat.\ Struct.\ Biol.\ }}

\newcommand  {\Phy}      {{\it Physica\ }}
\newcommand  {\Pro}      {{\it Proteins\ }}

\newcommand  {\ProSci}   {{\it Protein\ Sci.\ }}

\newcommand  {\PNAS}     {{\it Proc.\ Natl.\ Acad.\ Sci.\ USA\ }}

\newcommand  {\PRL}      {{\it Phys.\ Rev.\ Lett.\ }}

\newcommand  {\QRB}      {{\it Q.\ Rev.\ Biophys.\ }}  

\newcommand  {\Sci}      {{\it Science\ }}

\newcommand{\beq}{\begin{equation}}
\newcommand{\eeq}{\end{equation}}
\newcommand{\beqa}{\begin{eqnarray}}
\newcommand{\eeqa}{\end{eqnarray}}
\newcommand{\bea}{\begin{eqnarray}}
\newcommand{\eea}{\end{eqnarray}}
\newcommand   {\ev}[1]   {\langle #1\rangle}
\newcommand   {\Cb}      {C${}_{\beta}$}
\newcommand   {\dFU}     {\delta_{\mbox{\scriptsize FU}}} 
\newcommand   {\dBU}     {\delta_{\mbox{\scriptsize BU}}}

\newcommand   {\Xn}      {X_{\mbox{\scriptsize n}}} 
\newcommand   {\Xu}      {X_{\mbox{\scriptsize u}}} 
\newcommand   {\Pn}      {P_{\mbox{\scriptsize n}}} 
\newcommand   {\Pu}      {P_{\mbox{\scriptsize u}}} 
\newcommand   {\En}      {E_{\mbox{\scriptsize n}}} 
\newcommand   {\Eu}      {E_{\mbox{\scriptsize u}}} 

\newcommand   {\Ca}      {C${}_{\alpha}$}
\newcommand   {\Cp}      {C${}^{\prime}$}
\newcommand   {\dE}      {\Delta E}
\newcommand   {\dEsw}    {\Delta E_{\mbox{{\scriptsize sw}}}}
\newcommand   {\drsw}    {\Delta r_{\mbox{{\scriptsize sw}}}}
\newcommand   {\db}      {\delta \beta}
\newcommand   {\Eloc}    {E_{\mbox{{\scriptsize loc}}}}
\newcommand   {\Eev}     {E_{\mbox{{\scriptsize ev}}}}
\newcommand   {\Ehb}     {E_{\mbox{{\scriptsize hb}}}}
\newcommand   {\Ehp}     {E_{\mbox{{\scriptsize hp}}}}
\newcommand   {\Tm}      {T_{\mbox{{\scriptsize m}}}}
\newcommand   {\Tf}      {T_{\mbox{{\scriptsize f}}}}
\newcommand   {\Peq}     {P_{\mbox{{\scriptsize eq}}}}
\newcommand   {\tauf}    {\tau_{\mbox{{\scriptsize f}}}}
\newcommand   {\tauh}    {\tau_{\mbox{{\scriptsize h}}}}
\newcommand   {\taub}    {\tau_{\mbox{{\scriptsize b}}}}
\newcommand   {\taunaive}{\tau_{\mbox{{\scriptsize pred,0}}}}
\newcommand   {\taupred} {\tau_{\mbox{{\scriptsize pred}}}}
\newcommand   {\Chb}     {C_{\mbox{{\scriptsize hb}}}}

\newcommand   {\ephi}    {\epsilon_\phi}
\newcommand   {\epsi}    {\epsilon_\psi}
\newcommand   {\eev}     {\epsilon_{\mbox{{\scriptsize ev}}}}
\newcommand   {\ehb}     {\epsilon_{\mbox{{\scriptsize hb}}}}
\newcommand   {\ehp}     {\epsilon_{\mbox{{\scriptsize hp}}}}

\newcommand   {\shb}     {\sigma_{\mbox{{\scriptsize hb}}}}
\newcommand   {\shp}     {\sigma_{\mbox{{\scriptsize hp}}}}
\newcommand   {\Rg}      {R_{\mbox{{\scriptsize g}}}}
\newcommand   {\Db}      {D_{\mbox{{\scriptsize b}}}}
\newcommand   {\Fb}      {F_{\mbox{{\scriptsize b}}}}

\begin{document}

\begin{flushright}
LU TP 03-07\\
April 28, 2003
\end{flushright}

\vspace{0.4in}

\begin{center}

{\LARGE \bf Two-State Folding over a Weak Free-Energy Barrier}

\vspace{.6in}

\large
Giorgio Favrin, Anders Irb\"ack, Bj\"orn Samuelsson and Stefan 
Wallin\footnote{E-mail: favrin,\,anders,\,bjorn,\,stefan@thep.lu.se}\\   
\vspace{0.10in}
Complex Systems Division, Department of Theoretical Physics\\ 
Lund University,  S\"olvegatan 14A,  SE-223 62 Lund, Sweden \\
{\tt http://www.thep.lu.se/complex/}\\

\vspace{0.3in}	

\end{center}
\vspace{0.3in}
\normalsize
Abstract:\\
We present a Monte Carlo study of a model protein with 54 amino
acids that folds directly to its native three-helix-bundle state  
without forming any well-defined intermediate state. The free-energy
barrier separating the native and unfolded states of this protein
is found to be weak, even at the folding temperature. Nevertheless, 
we find that melting curves to a good approximation can be described 
in terms of a simple two-state system, and that the relaxation behavior  
is close to single exponential. The motion along
individual reaction coordinates is roughly diffusive on timescales beyond 
the reconfiguration time for an individual helix. A simple estimate
based on diffusion in a square-well potential predicts the relaxation  
time within a factor of two.


\newpage

\section{Introduction}

In a landmark paper in 1991, Jackson and Fersht~\cite{Jackson:91}  
demonstrated that chymotrypsin inhibitor 2 folds without 
significantly populating any meta-stable intermediate state. 
Since then, it has become clear that this protein is far from unique; 
the same behavior has been observed for many small 
single-domain proteins~\cite{Jackson:98}. It is tempting to interpret the 
apparent two-state behavior of these proteins in terms of a simple free-energy
landscape with two minima separated by a single barrier, where the minima
represent the native and unfolded states, respectively. If the barrier is 
high, this picture provides an explanation of why the folding kinetics 
are single exponential, and why the folding thermodynamics show two-state 
character.

However, it is well-known that the free-energy barrier, $\Delta F$, 
is not high for all these proteins. In fact, assuming the 
folding time $\tauf$ to be given by $\tauf=\tau_0\exp(\Delta F/kT)$ 
with $\tau_0\sim 1\,\mu s$~\cite{Hagen:96}, it is easy to find examples
of proteins with $\Delta F$ values of a few $kT$ ~\cite{Jackson:98}  
($k$ is Boltzmann's constant and $T$ the temperature). It should also be 
mentioned that Garcia-Mira~\etal~\cite{Garcia-Mira:02} recently found 
a protein that appears to fold without crossing any free-energy barrier.      

Suppose the native and unfolded states coexist at the folding temperature 
and that there is no well-defined intermediate state, but that a clear
free-energy barrier is missing. What type of folding behavior should
one then expect? In particular, would such a protein, due to the lack
of a clear free-energy barrier, show easily detectable deviations
from two-state thermodynamics and single-exponential kinetics? 
Here we investigate this question based on Monte Carlo 
simulations of a designed three-helix-bundle 
protein~\cite{Irback:00,Irback:01,Favrin:02}. 

Our study consists of three parts. First, we investigate whether 
or not melting curves for this model protein show two-state character. 
Second, we ask whether the relaxation behavior is single exponential 
or not, based on ensemble kinetics at the folding temperature. 
Third, inspired by energy-landscape theory (for a recent review, 
see Refs.~\cite{Plotkin:02a,Plotkin:02b}), we try to interpret the folding 
dynamics of this system in terms of simple diffusive motion 
in a low-dimensional free-energy landscape. 

\newpage

\section{Model and Methods}

\subsection{The Model}

Simulating atomic models for protein folding remains a challenge, 
although progress is currently being made in this 
area~\cite{Kussell:02,Shen:02,Zhou:02,Shea:02,
Zagrovic:02,Clementi:03,Irback:03}. 
Here, for computational efficiency, we consider a reduced model with 5--6 
atoms per amino acid~\cite{Irback:00}, in which the side chains are replaced 
by large \Cb\ atoms. Using this model, we study a designed 
three-helix-bundle protein with 54 amino acids.  
  
The model has the Ramachandran torsion angles $\phi_i$, $\psi_i$ as its 
degrees of freedom, and is sequence-based with three amino acid types:
hydrophobic (H), polar (P) and glycine (G). The sequence studied 
consists of three identical H/P segments with 16 amino acids
each (PPHPPHHPPHPPHHPP), separated by two short GGG 
segments~\cite{Guo:96,Takada:99}.
The H/P segment is such that it can make an $\alpha$-helix
with all the hydrophobic amino acids on the same side.     

The interaction potential 
\beq
E=\Eloc+\Eev+\Ehb+\Ehp
\label{e}\eeq
is composed of four terms.
The local potential $\Eloc$ has a standard form with threefold symmetry,
\beq
\Eloc=\frac{\ephi}{2}\sum_i(1 + \cos3\phi_i)
+ \frac{\epsi}{2}\sum_i(1 + \cos3\psi_i)\,.
\eeq
The excluded-volume term $\Eev$ is given by a hard-sphere potential
of the form
\beq
\Eev=\eev\mathop{{\sum}'}_{i<j}
\bigg(\frac{\sigma_{ij}}{r_{ij}}\bigg)^{12}\,,
\label{sa}\eeq
where the sum runs over all possible atom pairs except those
consisting of two hydrophobic \Cb. The parameter $\sigma_{ij}$
is given by $\sigma_{ij}=\sigma_i+\sigma_j+\Delta\sigma_{ij}$,
where $\Delta\sigma_{ij}=0.625$\,\AA\ for \Cb\Cp, \Cb N and \Cb O pairs
that are connected by a sequence of three covalent bonds, and 
$\Delta\sigma_{ij}=0$\,\AA\
otherwise. The introduction of the parameter $\Delta\sigma_{ij}$
can be thought of as a change of the local potential.
 
The hydrogen-bond term $\Ehb$ has the form
\beq
\Ehb= \ehb \sum_{ij}u(r_{ij})v(\alpha_{ij},\beta_{ij})\,,
\label{hb}\eeq
where the functions $u(r)$ and $v(\alpha,\beta)$ are given by
\begin{eqnarray}
u(r)&=&  5\bigg(\frac{\shb}{r}\bigg)^{12} -
        6\bigg(\frac{\shb}{r}\bigg)^{10}\label{u(r)}\\
v(\alpha,\beta)&=&\left\{
        \begin{array}{ll}
 \cos^2\alpha\cos^2\beta & \ \alpha,\beta>90^{\circ}\\
 0                      & \ \mbox{otherwise}
         \label{v(a,b)}\end{array} \right.
\end{eqnarray}
The sum in Eq.~\ref{hb} runs over all possible HO pairs, and
$r_{ij}$ denotes the HO distance, $\alpha_{ij}$ the NHO angle,
and $\beta_{ij}$ the HO\Cp\ angle. The last term of the potential,
the hydrophobicity term $\Ehp$, is given by
\beq
\Ehp=\ehp\sum_{i<j}\bigg[
\bigg(\frac{\shp}{r_{ij}}\bigg)^{12}
-2\bigg(\frac{\shp}{r_{ij}}\bigg)^6\,\bigg]\,,
\eeq
where the sum runs over all pairs of hydrophobic \Cb.

To speed up the calculations, a cutoff radius $r_c$ is used,
which is taken to be 4.5\,\AA\ for $\Eev$
and $\Ehb$, and 8\,\AA\ for $\Ehp$.
Numerical values of all energy and geometry parameters can be
found elsewhere~\cite{Irback:00}.

The thermodynamic behavior of this three-helix-bundle protein 
has been studied before~\cite{Irback:00,Irback:01}. These 
studies demonstrated that this model protein has the following 
properties:
\begin{itemize}
\item It does form a stable three-helix bundle, except for a
twofold topological degeneracy. These two topologically distinct 
states both contain three right-handed helices. They differ in how   
the helices are arranged. If we let the first two helices
form a U, then the third helix is in front of the U in one case (FU), and
behind the U in the other case (BU). The reason that the model is unable 
to discriminate between these two states is that their contact maps are 
effectively very similar~\cite{Wallin:03}.
\item It makes more stable helices than the corresponding one- and
two-helix sequences, which is in accord with the experimental fact
that tertiary interactions generally are needed for secondary structure 
to become stable.
\item It undergoes a first-order-like folding transition directly
from an expanded state to the three-helix-bundle state, 
without any detectable intermediate state. At the
folding temperature $\Tf$, there is a pronounced peak in the 
specific heat. 
\end{itemize}

Here we analyze the folding dynamics of this protein in more detail,
through an extended study of both thermodynamics and kinetics.
 
As a measure of structural similarity with the native state, we
monitor a parameter $Q$ that we call nativeness (the same
as in \cite{Irback:00, Irback:01,Favrin:02}). To calculate $Q$, we use
representative conformations for the FU and BU topologies,
respectively, obtained by energy minimization. For a given
conformation, we compute the root-mean-square deviations $\dFU$ and
$\dBU$ from these two representative conformations (calculated over
all backbone atoms). The nativeness $Q$ is then obtained as \beq
Q=\max\left[\exp\left(-\dFU^2/(10\mbox{\AA})^2\right),\,
\exp\left(-\dBU^2/(10\mbox{\AA})^2\right)\right]\,,
\label{Q}\eeq
which makes $Q$ a dimensionless number between 0 and 1.

Energies are quoted in units of $k\Tf$, with the folding temperature
$\Tf$ defined as the specific heat maximum. In the dimensionless 
energy unit used in our previous study~\cite{Irback:00}, this
temperature is given by $k\Tf=0.6585\pm0.0006$. 

\subsection{Monte Carlo Methods}

To simulate the thermodynamic behavior of this model, we use    
simulated tempering~\cite{Lyubartsev:92,Marinari:92,Irback:95},
in which the temperature is a dynamic variable. This method is chosen
in order to speed up the calculations at low temperatures. Our simulations   
are started from random configurations. The temperatures 
studied range from $0.95\,\Tf$ to $1.37\,\Tf$. 

The temperature update is a standard Metropolis step. 
In conformation space we use two different elementary moves: 
first, the pivot move in which a single 
torsion angle is turned; and second, a semi-local method~\cite{Favrin:01} 
that works with seven or eight adjacent torsion angles, which are turned 
in a coordinated manner. The non-local pivot move is included in our 
calculations in order to accelerate the evolution 
of the system at high temperatures.  

Our kinetic simulations are also Monte Carlo-based, and only 
meant to mimic the time evolution of the system in a qualitative 
sense. They differ from our thermodynamic simulations  
in two ways: first, the temperature is held constant; and second,
the non-local pivot update is not used, 
but only the semi-local method~\cite{Favrin:01}. 
This restriction is needed 
in order to avoid large unphysical deformations of the chain. 

Statistical errors on thermodynamic results are obtained by  
jackknife analysis~\cite{Miller:74} of results from ten or more 
independent runs, each containing several 
folding/unfolding events. All errors quoted are 
1$\sigma$ errors. 
Statistical errors on relaxation times are difficult
to determine due to uncertainties about where the large-time
behavior sets in and are therefore 
omitted. We estimate that the uncertainties on our calculated 
relaxation times are about 10\%. The statistical errors 
on the results obtained by numerical solution of the 
diffusion equation are, however, significantly smaller than this.  

All fits of data discussed 
below are carried out by using a Levenberg-Marquardt 
procedure~\cite{NR}.      

\subsection{Analysis}

Melting curves for proteins are often described in terms of a 
two-state picture. In the two-state approximation, the average 
of a quantity $X$ at temperature $T$ is given by 
\beq
X(T)=\frac{\Xu+\Xn K(T)}{1+K(T)}\,, 
\label{twostate}\eeq
where $K(T)=\Pn(T)/\Pu(T)$, $\Pn(T)$ and $\Pu(T)$ being the populations of the 
native and unfolded states, respectively. Likewise, $\Xn$ and $\Xu$
denote the respective values of $X$ in the native and unfolded states. 
The effective equilibrium constant $K(T)$ 
is to leading order given by $K(T)=\exp[(1/kT-1/k\Tm)\dE]$, 
where $\Tm$ is the midpoint temperature and $\dE$ the energy difference
between the two states. 
With this $K(T)$, a fit to Eq.~\ref{twostate} has four parameters: 
$\dE$, $\Tm$ and the two baselines $\Xu$ and $\Xn$.

A simple but powerful method for quantitative analysis of the folding
dynamics is obtained by assuming the motion along different 
reaction coordinates to be 
diffusive~\cite{Bryngelson:95,Socci:96}. The 
folding process is then modeled as one-dimensional
Brownian motion in an external potential given by the free energy 
$F(r)=-kT\ln\Peq(r)$, where $\Peq(r)$ denotes the equilibrium distribution
of $r$. Thus, it is assumed that the probability distribution of $r$ at 
time $t$, $P(r,t)$, obeys Smoluchowski's diffusion equation     
\beq
\frac{\partial P(r,t)}{\partial t}=\frac{\partial}{\partial r}\left 
[D(r)\left(\frac{\partial P(r,t)}{\partial r}+ 
\frac{P(r,t)}{kT}\frac{\partial F(r)}{\partial r}\right)\right]\,, 
\label{smol}\eeq
where $D(r)$ is the diffusion coefficient. 

This picture is not expected to hold on short timescales, 
due to the projection onto a single coordinate $r$, but may still be 
useful provided that the diffusive behavior sets in on a timescale that 
is small compared to the relaxation time. By estimating $D(r)$ and $F(r)$, 
it is then possible to predict the relaxation time from Eq.~\ref{smol}. 
Such an analysis has been successfully carried through 
for a lattice protein~\cite{Socci:96}. 

The relaxation behavior predicted by Eq.~\ref{smol} is well understood  
when $F(r)$ has the shape of a double well with a clear barrier. In this 
situation, the relaxation is single exponential with a rate constant given by 
Kramers' well-known result~\cite{Kramers:40}. However, this result cannot be
applied to our model, in which the free-energy barrier is small or absent, 
depending on which reaction coordinate is used. Therefore, we perform 
a detailed study of Eq.~\ref{smol} for some relevant choices of $D(r)$ and 
$F(r)$, using analytical as well as numerical methods.       

\section{Results}

\subsection{Thermodynamics} 

In our thermodynamic analysis, we study the five different
quantities listed in Table~\ref{tab:1}. The first question we ask is 
to what extent the temperature dependence of these quantities 
can be described in terms of a first-order two-state system 
(see Eq.~\ref{twostate}). 

\begin{table}[t]
\begin{center}
\begin{tabular}{lcc}
            &  $\dE/k\Tf$     & $\Tm/\Tf$ \\
\hline 

$E$         & 40.1 $\pm$ 3.3 & 1.0050 $\pm$ 0.0020 \\
$\Ehb$      & 41.0 $\pm$ 2.6 & 1.0024 $\pm$ 0.0017 \\
$\Ehp$      & 45.4 $\pm$ 3.3 & 1.0056 $\pm$ 0.0017 \\
$\Rg$       & 45.7 $\pm$ 3.8 & 1.0099 $\pm$ 0.0018 \\
$Q$         & 53.6 $\pm$ 2.1 & 0.9989 $\pm$ 0.0008 \\
\end{tabular}
\caption{Parameters $\dE$ and $\Tm$ obtained by fitting results 
from our thermodynamic simulations to the two-state expression
in Eq.~\protect\ref{twostate}. This is done individually for 
each of the quantities in the first column; 
the energy $E$,
the hydrogen-bond energy $\Ehb$, the hydrophobicity energy $\Ehp$,
the radius of gyration $\Rg$ (calculated over all backbone atoms), 
and the nativeness $Q$ (see Eq.~\ref{Q}). 
The fits are performed using seven data points in 
the temperature interval $0.95\,\Tf\leq T\leq 1.11\,\Tf$.}
\label{tab:1}
\end{center}
\end{table}

\begin{figure}[t]
\begin{center}
\mbox{
    \epsfig{figure=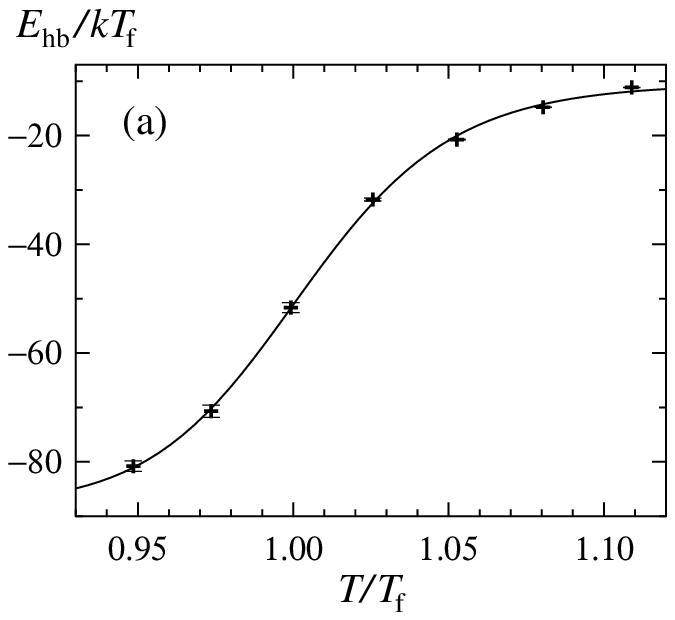}
\hspace{-3mm}
    \epsfig{figure=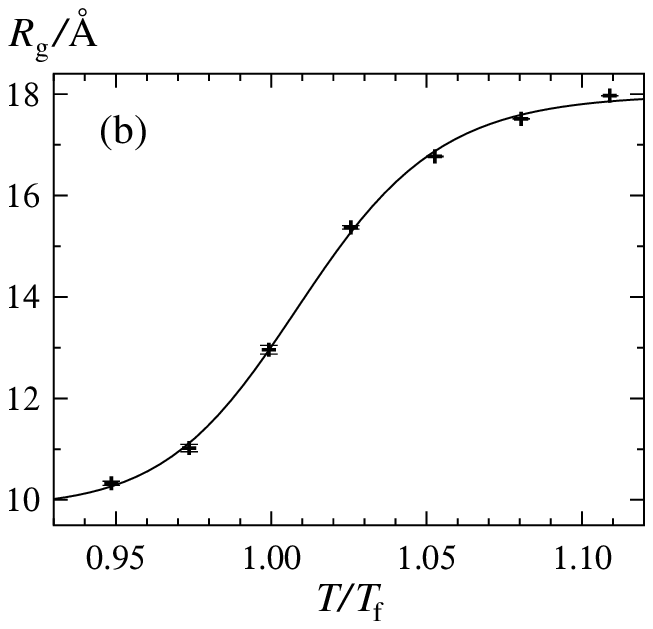}
}
\end{center}
\caption{Temperature dependence of (a) the hydrogen-bond energy  
$\Ehb$ and (b) the radius of gyration $\Rg$. The lines are fits
to Eq.~\ref{twostate}.}
\label{fig:1}
\end{figure}

Fits of our data to this equation show that the simple two-state
picture is not perfect 
($\chi^2$ per degree of freedom, dof, of $\sim$\,10), 
but this can
be detected only because the statistical errors are very small at high
temperatures ($<0.1\%$). In fact, if we assign artifical statistical
errors of 1\% to our data points, an error size that is not uncommon
for experimental data, then the fits become perfect with a
$\chi^2$/dof close to unity. Fig.~\ref{fig:1} shows the temperature
dependence of the hydrogen-bond energy $\Ehb$ and the radius of
gyration $\Rg$, along with our two-state fits.

Table~\ref{tab:1} gives a summary of our two-state fits.
In particular, we see that the fitted values of both the energy change $\dE$ 
and the midpoint temperature $\Tm$ are similar for the different quantities. 
It is also worth noting that the $\Tm$ values fall close to the folding 
temperature $\Tf$, defined as the maximum of the specific heat. 
The difference between the highest and lowest values 
of $\Tm$ is less than 1\%. There is a somewhat larger spread in 
$\dE$, but this parameter has a larger statistical error.  
   
\begin{figure}[h]
\begin{center}
\mbox{
  \mbox{
    \epsfig{figure=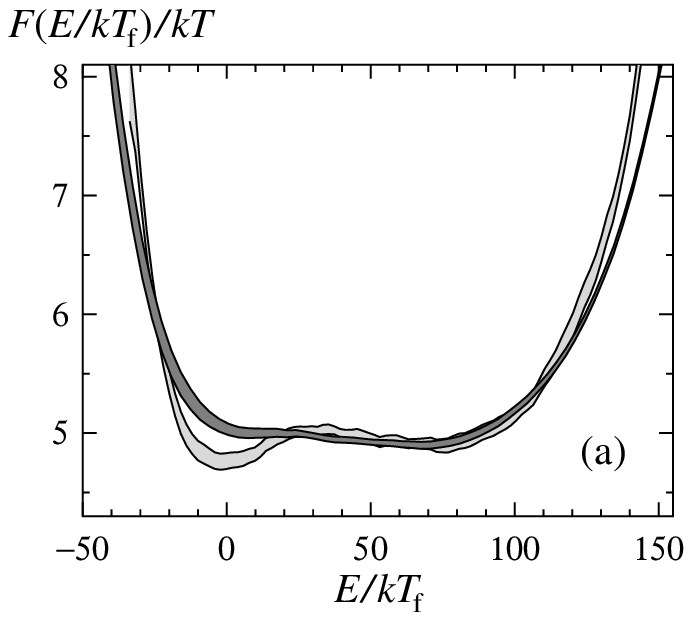}}
\hspace{-5mm}
  \mbox{
    \epsfig{figure=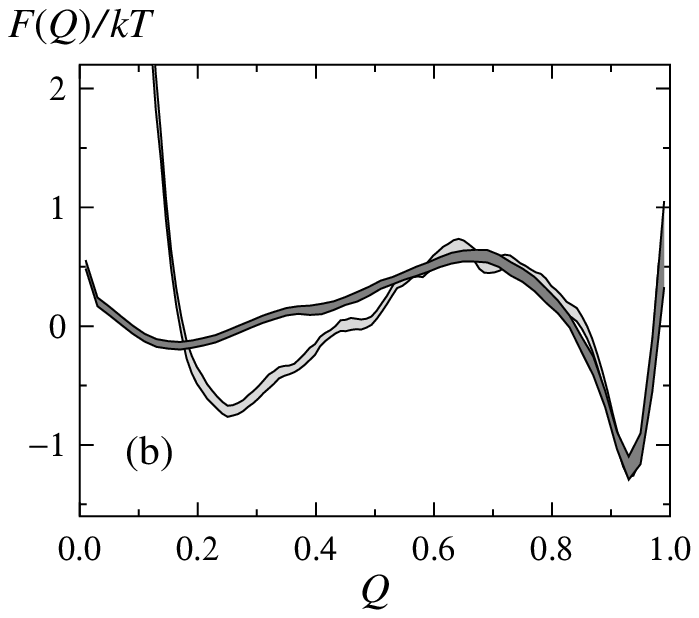}}
}
\end{center}
\caption{Free-energy profiles at $T=\Tf$ for (a) the energy $E$ and 
(b) the nativeness $Q$ (dark bands). The light-grey bands show 
free energies $\Fb$ for block averages (see Eq.~\ref{block}),
using a block size of $\taub=10^6$ MC steps. Each band is centered 
around the expected value and shows statistical 1$\sigma$ errors.}
\label{fig:2}
\end{figure}

So, the melting curves show two-state character, 
and the fitted parameters $\dE$ and $\Tm$ are similar for different 
quantities. From this it may be tempting to conclude that the 
thermodynamic behavior of this protein can be fully understood 
in terms of a two-state system. The two-state picture is, nevertheless, 
an oversimplification, as can be seen from the shapes of the 
free-energy profiles $F(E)$ and $F(Q)$. Fig.~\ref{fig:2} shows
these profiles at $T=\Tf$. First of all, these profiles show  
that the native and unfolded states coexist at $T=\Tf$, so the folding 
transition is first-order-like.  However, there is no clear free-energy 
barrier separating the two states; $F(Q)$ exhibits a very weak 
barrier, $<1\,kT$, whereas $F(E)$ shows no barrier at all. In fact, 
$F(E)$ has the shape of a square well rather than a double well.         

Phase transition terminology is, by necessity, ambiguous for a finite
system like this, but if states with markedly different $E$ or $Q$
coexist it does make sense to call the transition first-order-like,
even if a free-energy barrier is missing. At a second-order phase
transition, the free-energy profile is wide, but the minimum
remains unique.  

\subsection{Kinetics}

Our kinetic study is performed at $T=\Tf$. Using Monte Carlo
dynamics (see Model and Methods), we study the relaxation of ensemble 
averages of various quantities. For this purpose, 
we performed a set of 3000 folding simulations,
starting from equilibrium conformations at temperature
$T_0\approx1.06\,\Tf$. At this temperature, the chain is 
extended and has a relatively low secondary-structure content 
(see Fig.~\ref{fig:1}). 

\begin{figure}[t]
\begin{center}
\epsfig{figure=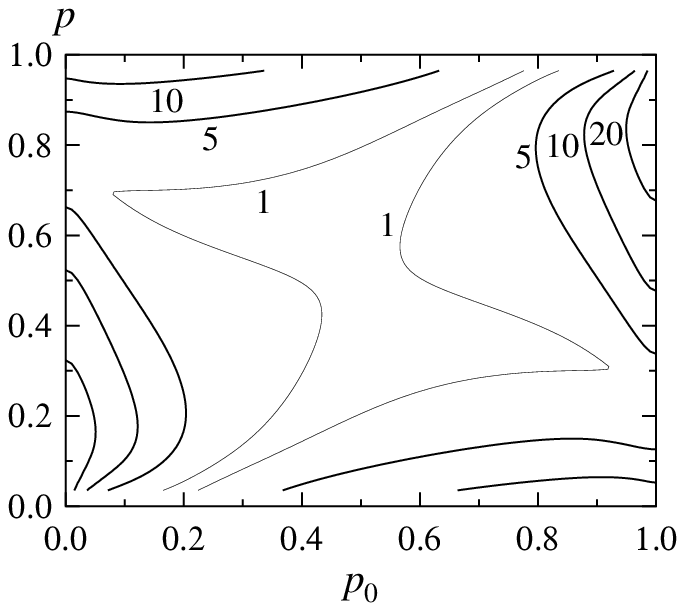}
\end{center}
\caption{Level diagram showing the deviation (in \%) from a single exponential 
for diffusion in energy in a square well, based on the exact solution in 
Appendix A. The system relaxes at temperature $T$, starting from the 
equilibrium distribution at temperature $T_0$.  
$p$ is defined as $p=(\ev{E}-\En)/\dEsw$, where $\ev{E}$ is the 
average energy at temperature $T$, and $\En$ and $\dEsw$ denote the 
lower edge and the width, respectively, of the square well.
$p$ can be viewed as a measure of the unfolded population
at temperature $T$, and is $0.5$ if $T=\Tf$. $p_0$ is the
the corresponding quantity at temperature $T_0$. As a measure of
the deviation from a single exponential, we take 
$\delta_{\max}/\delta E(t_0)$, where $\delta_{\max}$ is the
maximum deviation from a fitted exponential and 
$\delta E(t_0)=E(t_0)-\ev{E}$, $E(t_0)$ being the mean at the 
smallest time included in the fit, $t_0$.
Data at times shorter than 1\% of the 
relaxation time were excluded from the fit.} 
\label{fig:3}
\end{figure}

In the absence of a clear free-energy barrier (see Fig.~\ref{fig:2}),
it is not obvious whether or not the relaxation should be single
exponential.  To get an idea of what to expect for a system like this,
we consider the relaxation of the energy $E$ in a potential $F(E)$
that has the form of a perfect square well at $T=\Tf$. For this
idealized $F(E)$ and a constant diffusion coefficient $D(E)$,
it is possible to solve Eq.~\ref{smol} analytically
for relaxation at an arbitrary temperature $T$. This solution is given in
Appendix A, for the initial condition that $P(E,t=0)$ is the
equilibrium distribution at temperature $T_0$. Using this result, the
deviation from single-exponential behavior can be mapped out as a
function of $T_0$ and $T$, as is illustrated in Fig.~\ref{fig:3}. The
size of the deviation depends on both $T_0$ and $T$, but is found to
be small for a wide range of $T_0$, $T$ values. This clearly
demonstrates that the existence of a free-energy barrier is not a
prerequisite to observe single-exponential relaxation.

\begin{figure}[t]
\begin{center}
\mbox{
\epsfig{figure=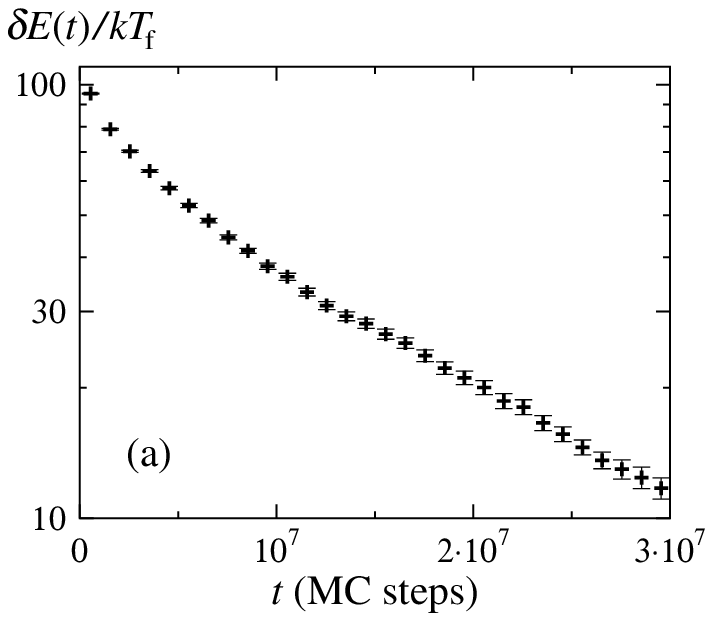}
\hspace{-5mm}
\epsfig{figure=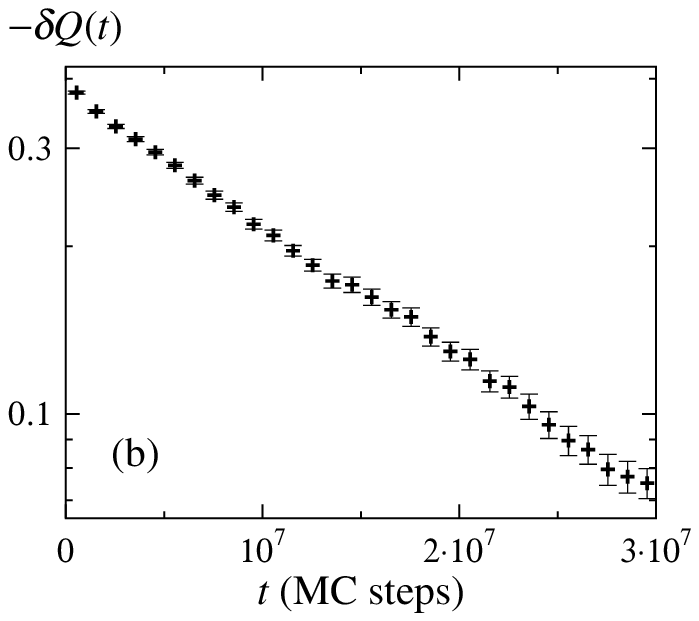}
}
\end{center}
\caption{Relaxation behavior of the three-helix-bundle protein at the
folding temperature $\Tf$, starting from the equilibrium ensemble at 
$T_0\approx 1.06\Tf$.
(a) $\delta E(t)=E(t)-\ev{E}$ against simulation time $t$, where $E(t)$ 
is the average $E$ after $t$ MC steps (3000 runs) and 
$\ev{E}$ denotes the equilibrium average (at $\Tf$).
(b) Same plot for the nativeness $Q$.} 
\label{fig:4}
\end{figure}

Let us now turn to the results of our simulations.
Fig.~\ref{fig:4} shows the relaxation of the average energy $E$ and 
the average nativeness $Q$ in Monte Carlo (MC) time. In both cases,
the large-time data can be fitted to a single exponential, which  
gives relaxation times 
of $\tau\approx 1.7\cdot 10^7$ and $\tau\approx 1.8\cdot 10^7$ 
for $E$ and $Q$, respectively, in units of elementary MC steps. 
The corresponding fits for the radius of gyration 
and the hydrogen-bond energy (data not shown) give relaxation 
times of $\tau\approx 2.1\cdot 10^7$ and $\tau\approx 1.8\cdot10^7$,
respectively. The fit for the radius of gyration has a larger 
uncertainty than the others, because the data points have larger 
errors in this case.   

The differences between our four fitted $\tau$ values are small and most 
probably due to limited statistics for the large-time behavior. 
Averaging over the four different variables, we obtain a relaxation time 
of $\tau\approx1.8\cdot 10^7$ MC steps for this protein. The fact that 
the relaxation times for the hydrogen-bond energy and the radius of gyration 
are approximately the same shows 
that helix formation and chain collapse proceed in parallel for this protein. 
This finding is in nice agreement with recent experimental results for small
helical proteins~\cite{Krantz:02}.

For $Q$, it is necessary to go to very short times
in order to see any significant deviation from a single exponential
(see Fig.~\ref{fig:4}). For $E$, we find that the single-exponential 
behavior sets in at roughly $\tau/3$, which means that the deviation 
from this behavior is larger than in the analytical calculation 
above. On the other hand, for comparisons with experimental data, 
we expect the behavior of $Q$ to be more relevant than that of $E$. 
The simulations confirm that the relaxation can be  
approximately single exponential even if there is no clear 
free-energy barrier. 

To translate the relaxation time for this protein into 
physical units, we compare with the reconfiguration time for the  
corresponding one-helix segment. To that end, we performed a 
kinetic simulation of this 16-amino acid segment at the same 
temperature, $T=\Tf$.
This temperature is above the midpoint temperature for the one-helix 
segment, which is $0.95\,\Tf$~\cite{Irback:00}. So, the isolated one-helix 
segment is unstable at $T=\Tf$, but makes frequent visits to
helical states with low hydrogen-bond energy, $\Ehb$. To obtain the
reconfiguration time, we fitted the large-time behavior of the
autocorrelation function for $\Ehb$, 
\beq
\Chb(t)=\ev{\Ehb(t)\Ehb(0)}-\ev{\Ehb(0)}^2\,, 
\label{autoc}\eeq 
to an exponential. The exponential autocorrelation time, which can be
viewed as a reconfiguration time, turned out to be  
$\tauh\approx1.0\cdot10^6$ MC steps. This is roughly a factor  
20 shorter than the relaxation time $\tau$ for the full three-helix 
bundle. Assuming the reconfiguration time for an individual 
helix to be $\sim$\,$0.2\,\mu$s~\cite{Williams:96,Thompson:97}, 
we obtain relaxation and folding times of $\sim$\,$4\,\mu$s 
and $\sim$\,$8\,\mu$s, respectively, for the three-helix bundle. 
This is fast but not inconceivable for a 
small helical protein~\cite{Jackson:98}. 
In fact, the B domain of staphylococcal protein A is a  
three-helix-bundle protein that has been found to fold in
$<10\,\mu$s, at 37${}^\circ$C~\cite{Myers:01}.

\subsection{Relaxation-Time Predictions}

We now turn to the question of whether the observed relaxation time 
can be predicted based on the diffusion equation, Eq.~\ref{smol}. 
For that purpose, we need to know not only the free energy $F(r)$, but 
also the diffusion coefficient $D(r)$. Socci~\etal~\cite{Socci:96} 
successfully performed this analysis for a lattice protein that 
exhibited a relatively clear free-energy barrier. Their estimate of $D(r)$ 
involved an autocorrelation time for the unfolded state. The absence
of a clear barrier separating the native and unfolded states makes it 
necessary to take a different approach in our case.

The one-dimensional diffusion picture is not expected to hold on short 
timescales, but only after coarse-graining in time. A computationally 
convenient way to implement this coarse-graining 
in time is to study block averages 
$b(t)$ defined by    
\beq
b(t)=\frac{1}{\taub}\sum_{t\le s< t+\taub}r(s)\qquad\qquad 
t=0, \taub, 2\taub,\ldots
\label{block}\eeq
where $\taub$ is the block size and $r$ is the reaction coordinate considered.
The diffusion coefficient can then be estimated 
using $\Db(r)=\ev{(\delta b)^2}/2\taub$, where the 
numerator is the mean-square difference between two consecutive
block averages, given that the first of them has the value $r$.  

In our calculations, we use a block size of $\taub=10^6$ MC step, 
corresponding to the reconfiguration time $\tauh$ for an 
individual helix. We do not expect the dynamics to be diffusive
on timescales shorter than this, due to steric traps that can occur
in the formation of a helix. In order for the dynamics to be diffusive,
the timescale should be such that the system can escape from 
these traps.   

Using this block size, we first make rough estimates of the
relaxation times for $E$ and $Q$ based on the result in Appendix A
for a square-well potential and a constant diffusion coefficient.
These estimates are given by 
$\taunaive=\drsw^2/\Db\pi^2$, where $\drsw$ is the width of 
the potential and $\Db$ is the average diffusion 
coefficient.~\footnote{Eq.~\protect\ref{rate} in Appendix A can be applied 
to other observables than $E$. The predicted relaxation time $\taunaive$ 
is given by $\tau_1$.} 
Our estimates of $\drsw$ and $\Db$
can be found in Table~\ref{tab:2}, along with the   
resulting predictions $\taunaive$. We find that these simple predictions
agree with the observed relaxation times $\tau$ within a 
factor of two.  

\begin{table}[t]
\begin{center}
\begin{tabular}{lccccc}
  & $\drsw$ & $\Db$ & $\taunaive$ & $\taupred$ & $\tau$ \\
\hline 
$E$: & $140k\Tf$ & $(9.3\pm0.2)\cdot10^{-5}(k\Tf)^2$ & $2.1\cdot10^7$ &   
$1.9\cdot10^7$ & $1.7\cdot10^7$\\
$Q$: & 1.0 & $(1.00\pm0.02)\cdot10^{-8}$ & $1.0\cdot10^7$ & 
$0.8\cdot10^7$ & $1.8\cdot10^7$
\end{tabular}
\caption{The predictions $\taunaive$ and $\taupred$ 
(see text) along with the observed relaxation time $\tau$,
as obtained from the data in Fig.~\protect\ref{fig:4}, for 
the energy $E$ and the nativeness $Q$. $\drsw$ is the width 
of the square-well potential and $\Db$ is the average diffusion 
coefficient.}
\label{tab:2}
\end{center}
\end{table}

We also did the same calculation for smaller block sizes, 
$\taub=10^0$, $10^1,\ldots$, $10^5$ MC steps. This gave  
$\taunaive$ values smaller or much smaller than the observed $\tau$,
signaling non-diffusive dynamics. This confirms that 
for $b(t)$ to show diffusive dynamics, $\taub$ should not be 
smaller than the reconfiguration time for an individual helix.   

\begin{figure}[t]
\begin{center}
\mbox{
    \epsfig{figure=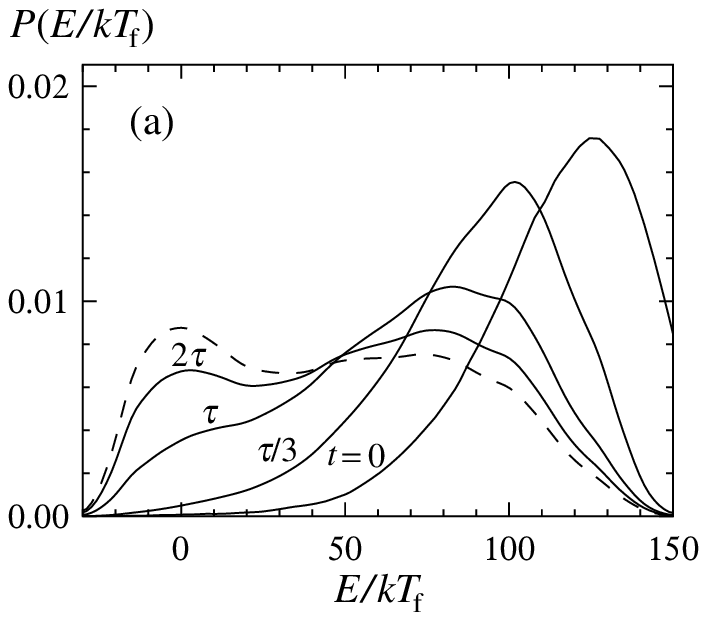}
\hspace{-5mm}
    \epsfig{figure=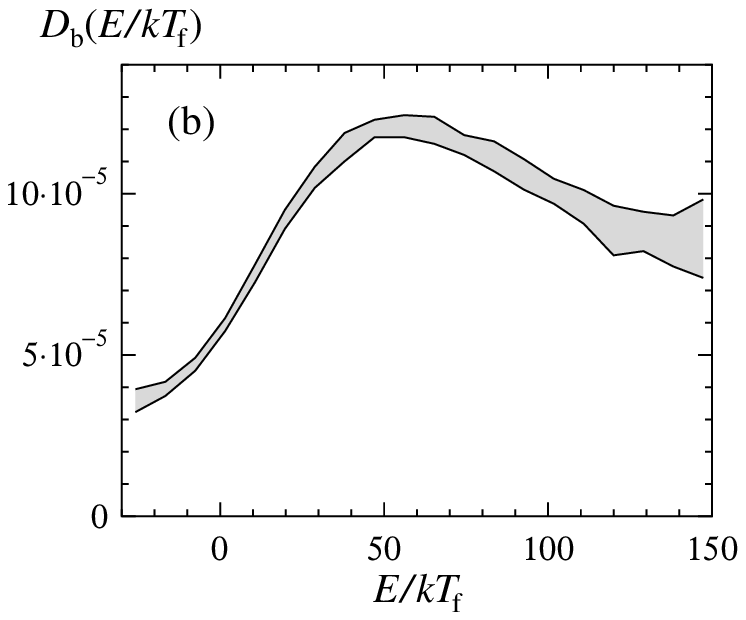}
}
\end{center}
\caption{(a) Numerical solution of Eq.~\ref{smol} with the energy as
reaction coordinate. The distribution $P(E,t)$ is shown  
for $t=0$, $\tau/3$, $\tau$ and $2\tau$ (full lines), where $\tau$ 
is the relaxation time. The dashed line is the equilibrium
distribution. The diffusion coefficient $\Db(E)$ and 
the potential $\Fb(E)$ (light-gray band in Fig.~\protect\ref{fig:2}a) 
were both determined from numerical simulations, using a block size of 
$\taub=10^6$ MC steps (see Eq.~\protect\ref{block}). (b) 
The space dependence of the diffusion coefficient $\Db(E)$.
The band is centered around the expected value and shows the
statistical $1\sigma$ error.}       
\label{fig:5}
\end{figure}

Having seen the quite good results obtained by this simple
calculation, we now turn to a more detailed analysis, illustrated in
Fig.~\ref{fig:5}a.  The block size is the same as before, $\taub=10^6$
MC steps, but the space dependence of the diffusion coefficient
$\Db(r)$ is now taken into account, and the potential, $\Fb(r)$,
reflects the actual distribution of block averages. 
The potential $\Fb(r)$, shown in Fig.~\ref{fig:2}, is not 
identical to that for the unblocked variables. At a first-order-like 
transition, we expect free-energy minima to become more pronounced 
when going to the block variables, provided that the block size 
$\taub$ is small compared with the relaxation time,   
because when forming the block variables one effectively integrates 
out fluctuations about the respective states. The results in
Fig.~\ref{fig:2} do show this tendency, although the effect is not 
very strong.    
Fig.~\ref{fig:5}b shows the diffusion
coefficient $\Db(E)$, which is largest at intermediate values between
the native and unfolded states.  The behavior of $\Db(Q)$ (not shown)
is the same in this respect.  Hence, there is no sign of a kinetic
bottleneck to folding for this protein.
     
Given $\Db(r)$ and $\Fb(r)$, we solve Eq.~\ref{smol} for $P(r,t)$ by using 
the finite-difference scheme in Appendix B. The initial distribution  
$P(r,t=0)$ is taken to be the same as in the kinetic simulations. 
We find that the mean of $P(r,t)$ shows single-exponential relaxation
to a good approximation. An exponential fit of these data gives us a new 
prediction, $\taupred$, for the relaxation time.      

From Table~\ref{tab:2} it can be seen that the prediction obtained
through this more elaborate calculation, $\taupred$, is not better
than the previous one, $\taunaive$, 
at least not in $Q$,
despite that there exists a weak barrier in this coordinate 
(see Fig.~2b). This means that the barrier in $Q$ is too weak 
to be important for the relaxation rate. If the underlying diffusion
picture, Eq.~\ref{smol}, had been perfect, $\taupred$ would have 
been equal to $\tau$, as obtained from the kinetic simulations.
Our results show that this is not the case. 
At least in $Q$, there are significant deviations from 
the behavior predicted by this equation. 
  
If more accurate relaxation time predictions are needed, 
there are different ways to proceed. One possible way is to simply
increase the block size. However, for the calculation to be useful, 
the block size must remain small compared to the relaxation time.
A more interesting possibility is to refine the simple diffusion 
picture defined by Eq.~\ref{smol}, in which, in particular, non-Markovian 
effects are ignored. Such effects may indeed affect 
folding times~\cite{Plotkin:98,Plotkin:02b}. 
Yet another possibility is to use a combination of a few different 
variables, perhaps $E$ and $Q$, instead of a single reaction 
coordinate~\cite{Du:97,Socci:98,Plotkin:02b}. With a multidimensional
representation of the folding process,  
non-Markovian effects could become smaller.   

\section{Summary and Discussion}

We have analyzed the thermodynamics and kinetics of a designed 
three-helix-bundle protein, based on Monte Carlo calculations. 
We found that this model protein shows two-state behavior, in the sense 
that melting curves to a good approximation can be described by 
a simple two-state system and that the relaxation behavior is close 
to single exponential. A simple two-state picture is, 
nevertheless, an oversimplification, as the free-energy barrier
separating the native and unfolded states is weak ($\lesssim1kT$). The 
weakness 
of the barrier implies that a fitted two-state parameter such as $\Delta E$ 
has no clear physical meaning, despite that the two-state fit looks 
good. 

Reduced~\cite{Kolinski:98,Takada:99,Zhou:99,Shea:99,Berriz:01} 
and all-atom~\cite{Guo:97,Duan:98,Shen:02,Kussell:02,Zagrovic:02,Linhananta:03}
models for small helical proteins have been studied by many other groups.
Most of these studies relied on so-called 
G\=o-type~\cite{Go:81} potentials. It should therefore be pointed out 
that our model is sequence-based.

Using an extended version of this model that includes all atoms,   
we recently found similar results for two peptides, an $\alpha$-helix 
and a $\beta$-hairpin~\cite{Irback:03}. Here the calculated melting 
curves could be directly compared with experimental data, and a reasonable 
quantitative agreement was found.  
 
The smallness of the free-energy barrier prompted us to perform
an analytical study of diffusion in a square-well potential. Here we 
studied the relaxation behavior at temperature $T$, starting from the 
equilibrium distribution at temperature $T_0$, for arbitrary $T$ 
and $T_0$. We found that this system shows a relaxation behavior that 
is close to single exponential for a wide range of $T_0$, $T$ values,
despite the absence of a free-energy barrier. We also made relaxation-time 
predictions based on this square-well approximation. Here we took the 
diffusion coefficient to be constant. It was determined assuming the 
dynamics to be diffusive on timescales beyond the 
reconfiguration time for an individual helix. The 
predictions obtained this way were found to agree within a factor of two
with observed relaxation times, as obtained from the kinetic simulations. 
So, this calculation, based on the two simplifying assumptions that
the potential is a square well and that the diffusion coefficient is constant, 
gave quite good results. A more detailed calculation, in which
these two additional assumptions were removed, did not give better
results. This shows that the underlying diffusion picture   
leaves room for improvement.        
   
Our kinetic study focused on the behavior at the folding temperature $\Tf$,
where the native and unfolded states, although not separated by a clear
barrier, are very different, which makes the folding mechanism transparent.
In particular, we found that 
helix formation and chain collapse could not be separated, 
which is in accord with experimental data by 
Krantz~\etal~\cite{Krantz:02}. The difference between the native 
and unfolded states is much smaller at the lowest temperature 
we studied, $0.95\Tf$, because the unfolded state is much more 
native-like here. Mayor~\etal~\cite{Mayor:03} recently reported 
experimental results on a three-helix-bundle protein,
the engrailed homeodomain~\cite{Clarke:94}, including a characterization 
of its unfolded state. In particular, the unfolded state was found to have 
a high helix content. This study was performed at a temperature below 
$0.95\Tf$. 
It would be very interesting to 
see what the unfolded state of this protein looks like near $\Tf$. 
In our model, there is a 
significant decrease in helix content of the unfolded state as the
temperature increases from $0.95\Tf$ to $\Tf$.    

It is instructive to compare our results with those of Zhou and
Karplus~\cite{Zhou:99}, who discussed two folding scenarios for
helical proteins, based on a G\=o-type \Ca\ model. In their first
scenario, folding is fast, without any obligatory intermediate, and
helix formation occurs before chain collapse. In the second scenario,
folding is slow with an obligatory intermediate on the folding
pathway, and helix formation and chain collapse occur simultaneously. 
The behavior we find does not match any of these two scenarios. 
In our case, helix formation and chain collapse occur in parallel
but folding is nevertheless fast and without any well-defined 
intermediate state.  
  
\subsection*{Acknowledgments}

This work was in part supported by the Swedish Foundation for Strategic
Research and the Swedish Research Council.

\newpage

\subsection*{Appendix A: Diffusion in a square well}

Here we discuss Eq.~\ref{smol} in the situation when the 
reaction coordinate $r$ is the energy $E$, and the potential $F(E)$ 
is a square well of width $\dEsw$ at $T=\Tf$. This means that the
equilibrium distribution is given by
$\Peq(E)\propto \exp(-\db E)$ if $E$ is in the square well and
$\Peq(E)=0$ otherwise, where $\db=1/kT-1/k\Tf$. Eq.~\ref{smol}
then becomes
\beq
  \frac{\partial P(E,t)}{\partial t} = \frac{\partial}{\partial E}
  \left[D\left(\frac{\partial P(E,t)}{\partial E} + \db P(E,t)\right)\right]\,.
\label{smole}\eeq
For simplicity, the diffusion coefficient is assumed 
to be constant, $D(E)=D$. The initial distribution $P(E,t=0)$ is taken   
to be the equilibrium distribution at some temperature $T_0$, and we put
$\db_0=1/kT_0-1/k\Tf$. 

By separation of variables, it is possible to solve Eq.~\ref{smole} 
with this initial condition analytically for arbitrary values of the
initial and final temperatures $T_0$ and $T$, respectively. In particular,
this solution gives us the relaxation behavior of the average energy. 
The average energy at time $t$, $E(t)$, can be expressed in the form     
\beq
  E(t) = \ev{E} + \sum_{k=1}^{\infty} A_k{\rm e}^{-t/\tau_k}\,, 
\eeq
where $\ev{E}$ denotes the equilibrium average at temperature $T$. A 
straightforward calculation shows that the decay constants in this
equation are given by 
\beq
  1/\tau_k = \frac{D}{\dEsw^2}
             \left(\pi^2k^2 + {\textstyle\frac14}\db^2\dEsw^2\right)
\label{rate}\eeq
and the expansion coefficients by
\beq
  A_k = B_k \dEsw 
            \frac{\pi^2k^2\left(\db-\db_0\right)\dEsw}
          {\left(\pi^2k^2 + (\db_0 - \frac12\db)^2\dEsw^2\right)
  \left(\pi^2k^2 + {\textstyle \frac14}\db^2\dEsw^2\right)^2}\,,
\eeq
where
\beq
  B_k=\frac{4\db_0\dEsw}{\sinh{\textstyle\frac12}\db_0\dEsw}\times
\left\{
\begin{array}{ll}
\cosh \left(\frac 12(\db_0 - \frac12\db) \dEsw \right)
            \cosh \frac 14 \db\dEsw &  {\rm if}\ k\ {\rm odd}\\  
\sinh \left(\frac 12(\db_0 - \frac12\db) \dEsw \right)
            \sinh \frac 14 \db\dEsw &  {\rm if}\ k\ {\rm even}  
\end{array}
\right.
\eeq
Finally, the equilibrium average is
\beq
  \ev{E} = \frac{\En + \Eu}2 + \frac{1}{\db} - \frac{\dEsw}2
          \coth{\textstyle \frac12}\db\dEsw\,, 
\label{eqE(sqwell)} \eeq
where $\En$ and $\Eu$ are the lower and upper edges of the 
square well, respectively.
 
It is instructive to consider the behavior of this solution
when $|\db-\db_0| \ll 1/\dEsw$. The expression for the expansion 
coefficients can then be simplified to 
\beq
  A_k \approx B_k\dEsw
          \frac{\pi^2k^2(\db-\db_0)\dEsw}
               {\left(\pi^2k^2 + \frac14\db^2\dEsw^2\right)^3}
\eeq
with
\beq
  B_k\approx\frac{4\db_0\dEsw}{\sinh{\textstyle\frac12}\db_0\dEsw}\times
\left\{
\begin{array}{ll}
            \cosh^2 \frac 14 \db\dEsw &  {\rm if}\ k\ {\rm odd}\\  
            \sinh^2 \frac 14 \db\dEsw &  {\rm if}\ k\ {\rm even}  
\end{array}
\right.
\eeq
Note that $A_k$ scales as $k^2$ if $k\ll\frac{1}{2\pi}|\db|\dEsw$, 
and as $1/k^4$ if $k\gg\frac{1}{2\pi}|\db|\dEsw$. Note also that
the last factor in $B_k$ suppresses $A_k$ for even $k$  
if $T$ is close to $\Tf$. From these two facts it follows that
$|A_1|$ is much larger than the other $|A_k|$ if $T$ is
near $\Tf$. This makes the deviation from a single exponential
small.  

\subsection*{Appendix B: Numerical solution of the diffusion equation}

To solve Eq.~\ref{smol} numerically for arbitrary $D(r)$ and
$F(r)$, we choose a finite-difference scheme of Crank-Nicolson type with
good stability properties. 
To obtain this scheme we first discretize $r$. Put $r_j=j\Delta r$,
$D_j=D(r_j)$ and $F_j=F(r_j)$, and let ${\bf p}(t)$ be the vector
with components $p_j(t)=P(r_j,t)$. Approximating the RHS of Eq.~\ref{smol}
with suitable finite differences, we obtain   
\beq
\frac{d{\bf p}}{dt}={\bf A}{\bf p}(t)\,,
\label{cna}\eeq
where {\bf A} is a tridiagonal matrix given by 
\begin{eqnarray}
({\bf A}{\bf p}(t))_j&=&
\frac{1}{\Delta r^2}
\left[D_{j+1/2}(p_{j+1}(t)-p_j(t))-D_{j-1/2}(p_j(t)-p_{j-1}(t))\right]
\nonumber\\
&+&\frac{1}{4kT\Delta r^2}
\left[D_{j+1}p_{j+1}(t)(F_{j+2}-F_j)-D_{j-1}p_{j-1}(t)(F_j-F_{j-2})\right]
\end{eqnarray}
Let now  ${\bf p}^n={\bf p}(t_n)$, where $t_n=n\Delta t$. By applying
the trapezoidal rule for integration to Eq.~\ref{cna}, we obtain
\beq
{\bf p}^{n+1}-{\bf p}^n=\frac{\Delta t}{2}
\left({\bf A}{\bf p}^n+{\bf A}{\bf p}^{n+1}\right)\,.
\eeq
This equation can be used to calculate how $P(r,t)$ evolves with time. 
It can be readily solved for ${\bf p}^{n+1}$ because the matrix 
${\bf A}$ is tridiagonal.

\end{document}